\documentclass[sigconf,nonacm]{acmart}
\setcopyright{none}

\usepackage{chapterbib}

\usepackage{url,hyperref,lineno,microtype,subcaption,verbatim}
\usepackage{booktabs}
\usepackage{tabularx}
\usepackage{colortbl}
\usepackage{adjustbox}

\usepackage[most]{tcolorbox}
\definecolor{block-gray}{gray}{0.87}
\newtcolorbox{citationblock}{colback=block-gray,
boxrule=0pt,boxsep=0pt,breakable}

\author{Karlijn Dinnissen}
\email{k.dinnissen@uu.nl}
\orcid{0000-0003-2498-2881}
\affiliation{
  \institution{Department of Information and Computing Sciences, Utrecht University}
  \city{Utrecht}
  \postcode{3584~CC}
  \streetaddress{Princetonplein~5}
  \country{The Netherlands}
}

\author{Christine Bauer}
\email{c.bauer@uu.nl}
\orcid{0000-0001-5724-1137}
\affiliation{
  \institution{Department of Information and Computing Sciences, Utrecht University}
  \city{Utrecht}
  \postcode{3584~CC}
  \streetaddress{Princetonplein~5}
  \country{The Netherlands}
}

\copyrightyear{2022} 
\acmYear{2022} 
\setcopyright{none}

\begin{document}

\title{A Stakeholder-Centered View on Fairness in Music Recommender Systems}

 \renewcommand{\abstractname}{Preamble}

\begin{abstract}
At the 5th FAccTRec Workshop on Responsible Recommendation (FAccTRec~'22)\footnote{\url{https://facctrec.github.io/facctrec2022/}}, we share and discuss the results of our recent literature review entitled ``Fairness in music recommender systems: a stakeholder-centered mini review'' published in Frontiers in Big Data.\footnote{Karlijn Dinnissen \& Christine Bauer. 2022. Fairness in Music Recommender Systems: A Stakeholder-Centered Mini Review. Frontiers in Big Data 5:913608. 9 pages. \url{https://doi.org/10.3389/fdata.2022.913608}}
\end{abstract}

\keywords{bias mitigation, fairness, music recommendation systems, stakeholders, literature review}


\maketitle

\begingroup
\let\clearpage\relax
\vspace{5mm}

\section*{FACCTREC~'22 ABSTRACT}

Our narrative literature review~\citep{dinnissen2022_fairness_in_music_recommender_systems} acknowledges that, although there is an increasing interest in recommender system fairness in general, the music domain has received relatively little attention in this regard. 
However, addressing fairness of music recommender systems (MRSs) is highly important because the performance of these systems considerably impacts both the users of music streaming platforms and the artists providing music to those platforms.
The distinct needs that these stakeholder groups may have, and the different aspects of fairness that therefore should be considered, make for a challenging research field with ample opportunities for improvement.
The review first outlines current literature on MRS fairness from the perspective of each stakeholder and the stakeholders combined, and then identifies promising directions for future research.

The two open questions arising from the review are as follows:
\begin{enumerate}
    \item In the MRS field, only limited data is publicly available to conduct fairness research; most datasets either originate from the same source or are proprietary (and, thus, not widely accessible).
    How can we address this limited data availability?
    
    \item Overall, the review shows that the large majority of works analyze the current situation of MRS fairness, whereas only few works propose approaches to improve it. How can we move forward to a focus on improving fairness aspects in these recommender systems?
\end{enumerate}

At FAccTRec~'22, we emphasize the specifics of addressing RS fairness in the music domain.
For example, a domain-specific challenge is the fact that the available music datasets are skewed.
This is, in part, caused by the historically grown imbalances with respect to, e.g., gender~\citep{schmutz2010_gender_cultural_consecration_music} and power relationships of major and indie labels~\citep{mall2018Concentration}.
These imbalances affect data quality and the biases that the available datasets encompass~\citep{zangerle2022_fevr}.
We explore to what extent solutions from other domains could be applied to tackle such fairness challenges and where novel, music domain-specific solutions are required.

\bibliographystyle{ACM-Reference-Format}
\bibliography{refs_fairness}

\vspace{40pt}

\begin{citationblock}
\vspace{5pt}
The remainder of this document is the accepted version of the paper appearing in Frontiers in Big Data.

\vspace{10pt}

\textit{Karlijn Dinnissen \& Christine Bauer. 2022. Fairness in Music Recommender Systems: A Stakeholder-Centered Mini Review. Frontiers in Big Data 5:913608. 9 pages. \url{https://doi.org/10.3389/fdata.2022.913608}
}
\vspace{5pt}

\end{citationblock}

\section*{Original abstract}

The performance of recommender systems highly impacts both music streaming platform users and the artists providing music.
As fairness is a fundamental value of human life, there is increasing pressure for these algorithmic decision-making processes to be fair as well.
However, many factors make recommender systems prone to biases, resulting in unfair outcomes.
Furthermore, several stakeholders are involved, who may all have distinct needs requiring different fairness considerations. 

While there is an increasing interest in research on recommender system fairness in general, 
the music domain has received relatively little attention.
This mini review, therefore, outlines current literature on music recommender system fairness from the perspective of each relevant stakeholder and the stakeholders combined.
For instance, various works address gender fairness: one line of research compares differences in recommendation quality across user gender groups, and another line focuses on the imbalanced representation of artist gender in the recommendations.
In addition to gender, popularity bias is frequently addressed; yet, primarily from the user perspective and rarely addressing how it impacts the representation of artists.
Overall, this narrative literature review shows that the large majority of works analyze the current situation of fairness in music recommender systems, whereas only a few works propose approaches to improve it.
This is, thus, a promising direction for future research.

\section{Introduction}
The art of music recommendation was traditionally performed exclusively by people, such as DJs, record store owners, and friends.
In the last few decades, however, this task has been partially automated using machine learning (ML) techniques; recommender systems (RSs) in particular~\citep{Celma2010_book}.
Learning from large-scale user behavior and music features, so-called music recommender systems (MRSs) can automatically produce recommendations tailored to a specific user~\citep{Ekstrand2022Fairness}.
This is one of the reasons why music streaming platforms, that typically integrate MRSs, have become one of the main sources of music consumption~\citep{IFPI2020}. 
Consequently, the performance of MRSs highly impacts users' overall music listening experience~\citep{lee2019can} and considerably impacts artists in terms of exposure and resulting royalty payments~\citep{Ferraro2021Fair}.

ML system users frequently perceive RS decisions as objective~\citep{HELBERGER2020105456}.
However, many factors make such systems' processes prone to biases, resulting in unfair outcomes~\citep{Ekstrand2022Fairness}.
One such factor is that ML models are created and trained by humans whose intrinsic biases may be carried over.
Furthermore, the data that is used to train ML models may contain biases as well.
This is problematic, as fairness is a fundamental value of human life~\citep{folger1998organizational,Tyler1998}.
Moreover, anti-discrimination regulations explicitly prohibit that characteristics such as gender, age, and nationality cause different outcomes for otherwise similar people~\citep[][Art. 21]{act1964civil, act1967age, EUCharter}.
It is, therefore, crucial to critically review MRSs for any form of unfairness to ensure that they do not unfairly disadvantage any user or artist.

Overall, there is an increasing interest in research on fairness in ML in general~\citep{Hutchinson2019Test}, and in RSs in particular~\citep{Ekstrand2019Fairness_recsys}. 
One of the challenges in fairness research is that it is scattered across several disciplines~\citep{Holstein2019,Selbst2019Fairness}.
Moreover, it concerns several stakeholders with distinct fairness needs, calling for various bias mitigation strategies~\citep{Ekstrand2022Fairness}.
Considering those needs is, thus, key to both understanding fairness in music recommendation algorithms and designing strategies to improve it.
To the best of our knowledge, an overview of such needs and strategies does not yet exist for the music recommendation field specifically.
Therefore, this work addresses the following research question: \emph{What is the state-of-the-art of MRS fairness research from the various stakeholders' perspectives?} 
To address this RQ, we conduct a narrative literature review, giving a thorough overview of works that \emph{explicitly} target \emph{RS fairness in the music domain}.
We also include some works that are not explicitly concerned with fairness, yet address fairness as a side effect. 

In Section~\ref{sec:multi}, we first define each relevant stakeholder group.
Then, in the Sections (\ref{sec:user}, \ref{sec:item}, and~\ref{sec:multistake}), we present our narrative literature review in which we address each of the relevant stakeholders separately.
In Section~\ref{sec:discussion}, we conclude this work with a discussion of the lessons learned from this overview and derive research gaps, thereby forming a solid basis for future research.

\section{Fairness for multiple stakeholders in music recommender systems}\label{sec:multi}

The digital music value chain embraces a wide set of stakeholders, who have different goals and interests regarding the music being recommended~\citep{bauer2019_multimethod_multistakeholder}.
Recommender systems literature typically distinguishes three stakeholders: platform users (end consumers), item providers, and the platform itself~\citep{Abdollahpouri2017Recommender, Burke2017Multisided, Sonboli2021Fairness}.
Some variations can be found in literature; for instance, \citet{Mehrotra2018Marketplace} and \citet{Patro2020Fairrec} only consider user and item provider as stakeholders, yet not the platform; 
conversely, \citet{Jannach2020Escaping} include society at large as a fourth stakeholder. 

In MRSs, there are three main stakeholders.
Firstly, the \textbf{users} (Section~\ref{sec:user})---also called consumers or customers---are the party consuming the music recommendations. 
A user may be an individual or a group of individuals, served by music streaming platforms.
As individuals have different profiles containing, for instance, different characteristics, preferences, or needs, MRSs might create a better experience for some user groups than for others.
Ideally, a MRS creates a good user experience for all users.

Secondly, the \textbf{item providers} (Section~\ref{sec:item})---also referred to as producers or suppliers---form the stakeholder supplying the recommended music and benefiting from it being consumed or purchased.
In MRS research, the artists (including performers, music producers, and songwriters) are typically the item providers, but record companies or publishers representing several artists may also be considered item providers.
Each item provider usually represents a multitude of items in the form of music tracks.
A higher MRS ranking for an item implies a higher chance of exposure to users, resulting in a higher chance that users interact with the item~\citep{Biega2018Equity, Diaz2020Evaluating}.
This is desirable, as item interaction results in revenue~\citep{Deldjoo2021Flexible}. 
Typically, item providers have little control over when and to whom their items are recommended~\citep{Burke2017Multisided,Ferraro2021Fair}.

Thirdly, the \textbf{platform} exists at the center of the music recommender ecosystem~\citep{Abdollahpouri2017Multiple, Smets2022Together}.
Music streaming platforms (such as Apple Music, Deezer, Pandora, QQ Music, Spotify, and Tidal) act as an interface between huge repositories of music tracks and millions of music consumers. On such platforms, the interaction between users and items is facilitated by a MRS. 
A platform needs to attract and retain users as well as item providers and, thus, benefits from a successful match between users and items~\citep{Burke2017Multisided}.
As the platforms are in control of the MRSs they embed~\citep{bauer2018_imbalance} and can even significantly influence consumption decisions through functionalities such as curated playlists~\citep{Aguiar2021Platforms}, they are typically not considered being at risk of unfair treatment. 
Rather, platforms might 
impose fairness constraints to satisfy an organizational mission or meet demands of, e.g., government regulators or interest groups~\citep{Ekstrand2022Fairness}.
Further, there is increasing external  
pressure to make these platforms and their integrated MRSs fairer~\citep{bauer2019_multimethod_multistakeholder,Burke2018Balanced,Ferraro2021Fair,Melchiorre2021Investigating,Patro2020Fairrec}. 

As multiple stakeholders with possibly diverging interests are involved and affected by MRSs, \textbf{multi-stakeholder} research (Section~\ref{sec:multistake}) addresses several stakeholder groups simultaneously.
Each stakeholder may have distinct fairness needs, which may further  
differ per context and application~\citep{Burke2017Multisided, Ekstrand2021Exploring}. 
Consequently, solely optimizing RSs on metrics such as user satisfaction may be detrimental to user fairness, item provider fairness, or both~\citep{bauer2019_multimethod_multistakeholder,Patro2020Fairrec}.
Hence, several studies urge to consider the interests of all stakeholder groups~\citep{Burke2017Multisided, Mehrotra2018Marketplace, Mehrotra2020Bandit}.
We note that research that addresses fairness, for example, for item providers, while also measuring performance indicators such as user satisfaction in the evaluation, are not necessarily multi-stakeholder approaches; a multi-stakeholder perspective integrates the various stakeholders fundamentally.

Table~\ref{tab:overview} provides an overview of the papers on fairness in MRSs considered in this narrative literature review. 
It also includes information on the research focus, methodology, considered fairness attributes, the stakeholders in the loop, and the datasets used for conducting the research.

\begin{table*}[htb]
    \caption{Overview of literature on fairness in music recommender systems.}
    \label{tab:overview}
    \footnotesize
    \centering
\begin{adjustbox}{totalheight=0.85\textheight-2\baselineskip}

\begin{minipage}{1.045\linewidth}
\begin{tabularx}{\textwidth}{>{\hsize=0.8cm}>{\raggedright\arraybackslash}X>{\hsize=1.1cm}>{\centering\arraybackslash}X>{\hsize=1.8cm}>{\raggedright\arraybackslash}X>{\raggedright\arraybackslash}X>{\hsize=2.8cm}>{\raggedright\arraybackslash}X>{\hsize=1.7cm}>{\raggedright\arraybackslash}X>{\hsize=1.4cm}>{\raggedright\arraybackslash}X}
\toprule
Reference & 
Improvement focus& 
Methodology
& 
Topic & 
Considered fairness attribute(s) & 
Stakeholder focus & 
Dataset source \\
\midrule
\arrayrulecolor{gray}\hline
\cite{Bauer2017Nonsuperstar}                                                       &                      & Conceptual, interview                                               & Negative impact for non-superstar artists                                                     & Popularity                                                                                     & Item provider                             & -- \\ 
\arrayrulecolor{gray}\hline
\cite{bauer2018_mainstreaminess_culture_maincult_hicss}                                                                        & x                    & Data analysis, offline experiment                                   & Improving accuracy by considering mainstreaminess and country                                 & User country, user `mainstreaminess'                                                             & User                                      & LFM-1b \\ \hline
\cite{bauer2019_plosone_mainstreaminess}                                                                        & x                    & Data analysis, offline experiment                                   & Improving accuracy by considering mainstreaminess and country                                 & User country, user `mainstreaminess'                                                             & User                                      & LFM-1b \\ \hline
\cite{Boratto2022Consumer}                                                                        &  x (repro-duction) 
& Systematic literature review, reproduction                                   & Reproducing and comparing unfairness mitigation strategies                                 & User age, user gender                                                             & User                                      & LFM-1K \\ \hline
\cite{Celma2010_book}                                                                                       &                      &  Data analysis 
& Promotion of niche items                                                                      & Popularity 
& User  
& Proprietary (Last.fm and MySpace)  
\\ \hline
\cite{Celma2008Hits}                                                                          &                      & Data analysis, offline experiment                                   & Investigating popularity bias in collaborative filtering                                      & Popularity 
& User 
& Proprietary (Last.fm)  
\\ \hline
\cite{Ekstrand2018Cool}   &                      & Data analysis, offline experiment                                   &             Recommender effectiveness across demographics and popularity level                                                                                  & Popularity, user age, user gender                                                              & User                                      & LFM-1K, LFM-360K \\ \hline
\cite{EppsDarling2020Gender}                                            &                      & Data analysis                                                       & Analysis of gender distribution across popularity levels                                      & Artist gender, popularity                                                                      & Item provider                             & Proprietary (Spotify) \\ \hline
\cite{Ferraro2020Artist}                                                           &                     & Offline experiment & Evaluating influence of recommendation bias on artist exposure                                                                  & Contemporaneity, country, gender, type (all artist attributes) & Item provider & LFM-360K \\ \hline
\cite{Ferraro2021Break}                                                           & x                    & Interviews, data analysis, offline experiment, long-term simulation & Improving gender fairness                                                                     & Artist gender & Item provider & LFM-360K, LFM-1b \\ \hline
\cite{Ferraro2021Fair} &                      & Interviews                                                          & Impact of recommender systems on artists                                                      & Age, contemporaneity, country, diversity, gender, popularity (all artist attributes)                      & Item provider & -- \\ \hline
\cite{Flexer2018Hubness}                                          &                      & Data analysis                                                       & Hubness as a technical algorithmic bias in high dimensional machine learning                  & 
--\footnote{Hubness can create unfairness for any attribute.}                   & User, item provider                       & Proprietary (FM4 SoundPark) 
\\ \hline
\cite{Htun2021Perception}                                                       & & User study & Perception of fairness per user personality type                                              & 
--\footnote{Not transparent which fairness attributes participants were considering.} & User & -- \\ \hline
\cite{kowald2021_support} & & Data analysis, offline experiment                                                       & Characteristics of niche music and music listeners                                            & User `mainstreaminess' & User & LFM-1b \\ \hline
\cite{Kowald2020Unfairness}                                                             &                      & Data analysis, offline experiment                                   & Investigating the impact of popularity bias on niche items, and users favoring those items   & Popularity, user `mainstreaminess'                                                            & User & LFM-1b \\ \hline
\cite{Lesota2021Analyzing}             &                      & Data analysis, offline experiment                                   & Effect of popularity bias per gender                                                          & Popularity, user gender 
& User & LFM-2b \\ \hline
\cite{Mehrotra2018Marketplace}                           & x & Offline experiment                                                  & Relevance, fairness and satisfaction trade-off in a two sided marketplace                     & Popularity                                                                                     & User, item provider                       & Proprietary (Spotify) \\ \hline
\cite{Mehrotra2020Bandit}                                                        & x                    & Offline experiment                                                  & Contextual bandits that consider multiple objectives (e.g., gender diversity, niche items)    & Artist gender, popularity                                                                      & User, item provider                       & Proprietary (Spotify), Simulated data \\ \hline
\cite{Melchiorre2021Investigating} & x                    & Data analysis, offline experiment                                   & Improvement of gender fairness considering popularity bias                                    & User gender & User & LFM-2b \\ \hline
\cite{Mousavifar2022}                                                             &                      & User study                                                          & Using explanations to increase user satisfaction with fair recommendation                     & Popularity                                                                                     & User, item provider                       & -- \\ \hline
\cite{Neophytou2022Revisiting}                                                             &                      & Offline experiment, reproduction                                                          & Reproducing recommendation utility for different user groups                     & Popularity, user age, user country, user gender                                                                                     & User                       & LFM-360K \\ \hline
\cite{Oliveira2017Multiobjective}                                 & x & Offline experiment                                                  & Considering diversification and user preferences simultaneously in a multi-objective approach & Contemporaneity, gender, genre, locality (all artist attributes)                                          & User, item provider                       & LFM-1b, Simulated data \\ \hline
\cite{schedl2017_distance_rank}                                                                        &                      & Offline experiment                                                  & Improving accuracy by considering mainstreaminess                                             & User `mainstreaminess'                                                                         & User                                      & LFM-1b \\ \hline
\cite{Shakespeare2020Exploring} &  & Data analysis, offline experiment                                   & Investigating gender fairness & Artist gender & Item provider & LFM-360K, LFM-1b, Simulated data \\ \arrayrulecolor{black}
\bottomrule
\end{tabularx}
\end{minipage}
\end{adjustbox}
\end{table*}

\normalsize

\subsection{User Perspective}\label{sec:user}

From the user perspective, fairness in MRSs is primarily studied based on distinct user groups defined by personal characteristics.
In addition to groups based on protected characteristics, groups differentiated by other characteristics may experience unfairness as well.

A wealth of literature analyzes popularity bias and subsequent mitigation strategies in various application domains \citep[e.g.,][]{Abdollahpouri20217Controlling,Figueiredo2014Youtube,Wei2021Model}.
It is, for instance, widely acknowledged that collaborative filtering-based recommendation approaches are prone to popularity bias~\citep{Celma2008Hits,Jannach2015What}. 
The music domain is a well-known example of the long-tail economy~\citep{Anderson2006_book} and popularity bias is, thus, particularly relevant.
It can be considered either a problem~\citep{Anderson2006_book} or a desired feature, as popularity in the community signifies some relevancy~\citep{Celma2010_book}.
In general, many works address popularity bias in MRSs with various intentions.
Some address the cold-start problem for items without prior user ratings to make them recommendable \citep[e.g.,][]{Ferraro2019Music}; others aim at increasing user satisfaction by adding novelty through recommending items from the long tail \citep[e.g.,][]{Bedi2014Using}; yet other works leverage the long tail to specifically address discovery \citep[e.g.,][]{Domingues2013Combining}.
While fairness is not always necessarily put in the loop of the investigation, this research thread does address fairness aspects.

As for insights from works that explicitly consider user fairness in MRSs, recommendation accuracy
tends to be higher for `mainstream' users, who are inclined toward what is popular, compared to `beyond-mainstream' users who prefer less popular items~\citep{kowald2021_support,Kowald2020Unfairness}. 
This also holds when defining user groups based on a more fine-grained music taste level~\citep{kowald2021_support,schedl2017_distance_rank}.
Some works \citep[e.g.,][]{bauer2019_plosone_mainstreaminess} have proposed mechanisms that better reflect the preferences of beyond-mainstream users. 
When defining user groups based on user country, popularity bias also negatively affects MRS performance for groups from countries with preferences beyond the global mainstream~\citep{bauer2018_mainstreaminess_culture_maincult_hicss,Neophytou2022Revisiting}.
In a later work, \citet{bauer2019_plosone_mainstreaminess} propose context-prefiltering approaches to mitigate this issue.

Zooming in on another user characteristic, several studies investigate gender. 
They show that popularity bias particularly affects minority gender groups (in these studies: women), resulting in lower-quality recommendations in terms of accuracy and coverage~\citep[e.g.,][]{Lesota2021Analyzing, Melchiorre2021Investigating}. 
In addition to finding similar results for user gender, \citet{Ekstrand2018Cool} and its reproducibility study by~\citet{Neophytou2022Revisiting} found performance differences for different user age groups, too. 
Here, the older user group received lower-quality recommendations. 

Lastly, on the mitigation side, \citet{Boratto2022Consumer} present a reproducibility study focusing on user age and gender, applying various mitigation strategies in the music and movie domains.
Different from the movie domain, the size of the user group was not indicative of the recommender accuracy in the music domain. Given their indecisive results, it is important to look beyond popularity bias and demographic group size to understand the drivers of demographic differences.

\citet{Melchiorre2020Personality} define user groups based on personality traits.
In contrast to the work on gender, age, and country, personality traits are not among the characteristics acknowledged by anti-discrimination regulations, and fairness research is not clear about this issue either. 
Nonetheless, they may be a source of bias and an opportunity for MRS improvement.
\citet{Melchiorre2020Personality} illustrate this by showing that scoring low on the personality traits openness, extraversion, and conscientiousness results in higher recommender performance, whereas scoring low on neuroticism or agreeableness leads to lower performance.
Additionally, \citet{Htun2021Perception} study the effect of personality traits on the perception of fairness in group recommendations when creating group music playlists.
Here, the personality trait openness is negatively correlated with the perception that fairness is important in groups.
Given that diversity needs and personality traits correlate~\citep{Chen2013How}, considering those traits in user modeling may help improve MRS performance.

\subsection{Item Provider Perspective}\label{sec:item}

When considering harm against music providers caused by unfairness in MRSs, research mainly focuses on group fairness~\citep{Singh2018Fairness}. 
Item provider groups in MRS research have been primarily defined based on gender~\citep{Ekstrand2021Exploring, Ferraro2021Break}. 
Several approaches are used to study and mitigate item provider gender bias, illustrating that a multifaceted approach is needed.
To date, most research has focused on understanding existing gender biases~\citep[e.g.,][]{EppsDarling2020Gender,Wang2019Gender}.
The former analyzed a Spotify streaming sample and found a disparity between artist genders in users' listening behavior.
In `organic' streaming, such as streams originating from a user library or user's search, 21.75\% of tracks were from either a woman or multi-gender formation.
For streams programmed by MRSs, this number was 23.55\%. 
This gender gap in listening behavior is further reflected in commonly used datasets such as LFM-1b and LFM-360k, in which 23\% of (solo) artists are women~\citep{Ferraro2021Break}.
These datasets roughly reflect the gender gap in business reality~\citep{EppsDarling2020Gender,Youngs2019}.
Overall, these percentages reflect the barriers to entry, and subsequently climbing to the top, for minority genders. 
In addition, pre-existing gender biases might influence which tracks users select in a MRS.
\citet{Ferraro2020Artist} and \citet{Shakespeare2020Exploring} found that collaborative filtering algorithms could propagate or even amplify those biases in a MRS, thereby negatively impacting minority genders.
In the latter, no evidence was found for the algorithms introducing new gender biases, which is supported by \citet{EppsDarling2020Gender} who found that recommendation-based streaming even contained  
a slightly higher proportion of tracks by women than in organic listening.
On the gender bias mitigation side, re-ranking is a promising method.
\citet{Ferraro2021Break} demonstrate breaking bias amplification through gradually increasing exposure for minority genders. 

In addition to gender, \citet{Oliveira2017Multiobjective} consider genre, locality, and contemporaneity. Embracing these attributes, they introduce a multi-objective approach to diversification that addresses fairness for users and item providers alike.
\citet{Ferraro2020Artist} use similar categories and add artist type (e.g., solo artist, band). 
Their analysis of the locality attribute indicates that group size may foster exposure: the artists from the most represented countries in the dataset (here: United Kingdom and United States) reached high exposure, while minority countries were penalized.

Defining item provider groups based on their popularity level has been investigated, too~\citep{Bauer2017Nonsuperstar, Celma2008Hits}.
Although popularity bias is a frequently researched topic, 
fairness goals are predominantly defined for MRS users and not item providers. 
One exception to this is \citet{Flexer2018Hubness} who study the `hubness' phenomenon, which 
can occur in content-based RS models that use song similarity as their main feature.
Hubness refers to some music tracks being connected to many other tracks in the database without a clear semantic musical connection.
This may introduce unfairness for tracks that are more similar semantically, but not recommended as often.

To date, one study directly discusses fairness in MRSs with the item providers themselves:
\citet{Ferraro2021Fair} interviewed artists about their perception of fairness in MRSs, and how item provider fairness could be improved on music streaming platforms. 
In those interviews, the main noted fairness improvement areas relate to nurturing diversity in general, and in particular to gender representation, addressing popularity bias, and providing a better representation of genres beyond the mainstream. 
These topics also correspond to the aforementioned research focuses in literature.

\subsection{Multi-Stakeholder Perspective}\label{sec:multistake}
Studies may simultaneously take several different MRS stakeholder objectives (e.g., satisfaction, utility, fairness, or diversity) into account.
Generally, across application domains, a trade-off between such objectives is reported~\citep{Cramer2018Assessing, Mehrotra2018Marketplace, Singh2018Fairness}, though it is possible that multi-stakeholder objective optimization benefits all stakeholders. 
Item provider fairness, for example, does not have to be detrimental to user satisfaction~\citep{Mehrotra2018Marketplace}, and persuasive strategies may even be implemented to promote new and less popular artists while increasing user satisfaction~\citep{Mousavifar2022}.
Furthermore, even if users do not directly benefit from or even consider fairness for item providers, they indicate that it is important to incorporate it in RSs~\citep{Sonboli2021Fairness}.

Overall, fairness-related multi-stakeholder MRS work mainly defines objectives and stakeholders rather than aiming to improve fairness.
\citet{Mehrotra2018Marketplace}, though, do contribute to fairness improvement by introducing 
a counterfactual estimation framework that balances provider fairness with user relevance and can optimize either, aiming to provide an alternative for expensive online A/B tests.
In another study, \citet{Mehrotra2020Bandit} use `contextual bandits' that can optimize multiple objectives simultaneously in a fair way, this time focusing on user- and platform objectives as opposed to item providers'. 

We might also draw inspiration from multi-stakeholder MRS research where fairness is not an explicitly defined goal.
For instance, \citet{Unger2021Deep} introduce a multi-objective RS that aims to fulfill both user satisfaction (measured by saves, likes, and engagement) and item provider satisfaction (determined by, e.g., acquiring new fans). 
A similar approach may be taken to implement fairness objectives for multiple stakeholders. 
\citet{Patro2020Fairrec} propose FairRec, which exhibits fairness for both user and item provider while the loss in overall recommendation quality remains marginal. FairRec has, however, not been applied to the music domain yet.

\section{Discussion and Conclusions}\label{sec:discussion}

This literature overview demonstrates that, while there is increasing interest in research on fairness in RSs in general, comparatively little research has addressed the music domain.
Below, we discuss the main findings we derive from this review.

\subsection{Research focus}
Contrary to what literature frequently claims~\citep[e.g.,][]{Ferraro2021Fair,Patro2020Fairrec}, 
fairness in this context has been addressed from both the user perspective and the item provider perspective.
Yet, multi-stakeholder approaches to fairness are scarce. 
This review also shows that the large majority of MRS fairness works analyze the current situation, using existing approaches and available datasets.
We, therefore, identify improvement-focused research as the main research gap.
A major challenge remains here: we still need to improve our understanding of the normative nature of fairness.
While an entirely fair system is likely unachievable, it is crucial to recognize RS fairness issues, mitigate them, and incrementally improve fairness over the current state.

\subsection{Gender bias}
Interestingly, various MRS works address gender fairness, both for user and item providers. 
We speculate that this focus has emerged from gender being an immutable characteristic, the wide acknowledgment that gender fairness is of societal relevance, and gender labels being available to some extent in relevant datasets. 
While it is a known limitation that a binary concept of gender oversimplifies gender expression, current datasets predominantly restrict the gender labels to man and woman~\citep{Boratto2022Consumer,Ferraro2021Break,Shakespeare2020Exploring}.
A notable exception is the work by \citet{EppsDarling2020Gender}.

\subsection{Popularity bias}
While popularity bias may be considered an item provider fairness issue as the gap between popular and unpopular items increases, research frequently focuses on the user.
Addressing popularity is seen as a means to provide more diverse content to increase user satisfaction. 
Similarly, we observe that some works do not explicitly focus on fairness, but still demonstrate fairness intentions or improvements in their research.
As this review focused on works that address fairness \emph{explicitly}, this overview is not intended to be exhaustive. 

\subsection{Data availability}
As can be seen in Table~\ref{tab:overview}, the most frequently used datasets originate from Last.fm: LFM-1b \citep{Schedl2016LFM}, LFM-1K, LFM-360K \citep[both][]{Celma2010Music}, and the recently added LFM-2b \citep{Schedl2022LFM}.
This results in only a few datasets being used for research on fairness in MRSs; most of which are either based on the same or similar Last.fm data, or are proprietary and therefore not accessible to other researchers.
Overall, this means that the used datasets might not be representative. 
Additionally, only a few open datasets in the music domain contain user interaction or preference data. 
They also typically include only limited fairness-related stakeholder metadata (e.g., gender, age, ethnicity), as sensitive data is often not shared~\citep{Stoikov2021Evaluating}. 
For ethical reasons, it is debatable whether it should be.
Lastly, a current limitation is the focus on short-term bias mitigation, while real world-systems are active over years~\citep{Shakespeare2020Exploring}.
Longitudinal data or simulation frameworks are needed to better address these temporary aspects and to study fairness in MRSs in the long run.
Summing up, to achieve significant MRS fairness improvements, richer and more representative data is needed.


\section*{Author Contributions}
KD and CB contributed to writing and revising the manuscript draft, as well as the final submitted version.

\bibliographystyle{ACM-Reference-Format}
\bibliography{refs_fairness}

\endgroup

\end{document}